# On Concept of Petri Nets' Receptors and Effectors

**Alexander Yu. Chunikhin,**
**Marina D. Sviatnenko**
Palladin Institute of Biochemistry
National Academy of Sciences of Ukraine
alexchunikhin61@gmail.com
marinasviatnenko0202@gmail.com

**Abstract**. New subclasses of Petri nets - Petri nets' receptors and Petri nets' effectors are introduced. The introduction/exclusion of such substructures in the main Petri net may be fulfilled in accordance with the Fusion/Defusion principles. We propose two pairs of entities: position marking receptor (effector) and transition marking receptor (effector), which allow to observe parameters of the main Petri net and, if necessary, to carry out their regulation.

## 1. Introduction and Motivation

First introduced in 1962 [15], Petri nets, without exaggeration, have become, along with artificial neural networks, one of the main technologies for modeling complex dynamic systems in various fields of human practice. Along with discrete and continuous Petri nets, hybrid (discrete-continuous) [5] and hybrid-functional [10] implementations of Petri nets are also being developed. Among the various extensions of Petri nets, the most common in applications are Timed [16, 17], Colored [6], Stochastic [4], Fuzzy [1, 8] and Rough [11, 13] Petri nets. Recently, such types of networks as Nested [9], Open [3] and Reconfigurable [12] Petri nets are actively developing.

At the same time, most real systems are complex systems, for the effective functioning of which both a channel for monitoring (measuring) the state of the system and a channel for controlling (directed exposure) are needed. In the vast majority of Petri nets models, these channels either remain "behind the scenes" (by default) or are included in the structure of the original model as an integral part. For holistic system modeling, this approach seems unsatisfactory, since it does not reflect many aspects of monitoring and controlling the system, namely: types of state observers (sensors, receptors) and effectors (actuators), places (points, positions) of their connection, moments of connection/disconnection, their presence during the entire cycle of the system or only at certain intervals, processing the received data about the network state and making a decision on the control vector, etc.

It is proposed to separate the measuring and the controlling structures into corresponding subclasses of Petri nets, in particular, Petri nets' receptors (PN-receptors) and Petri nets' effectors (PN-effectors), respectively. The introduction/exclusion of such substructures in the structure of the Petri net model during the process of its functioning (performance) becomes possible owing to the Fusion/Defusion Principles introduced in [2].

For the teamwork of receptors and effectors we introduced one more complex structure, called *Solver*. The purpose of this structure is to analyze received information from the receptors and determine the control vector for the effectors.

## 2. Petri Nets' Structural Units and Fusion/ Defusion Principles [2]

Components of the Petri net are traditionally considered as positions, transitions and arcs [14]. However, neither the components themselves (O, →, |) nor any of their permissible paired formations (O→, →|, |→, →O) have a complete sense in terms of the performing elementary operations in the Petri net. Rather, positions, transitions and arcs can be called *proto-elements* of the Petri net.

**Proposition 1.** The *structural units* of the Petri net are defined as the "position-arc-transition" triad: O→| and the "transition-arc-position" triad: |→O.

They are the basic comprehensive structural formations, "bricks", on the basis of which the structure of an arbitrary Petri net is constructed and varied explicitly (without hidden assumptions).

**Proposition 2.** *Fusion Principle*. Any elementary Petri net can be designed from structural units (p-units and t-units) using fusion operators of two types: *p-fusion* is n-ary operator of composition of structural units by fusing their positions pf(_, ..., _); *t-fusion* is an n-ary operator of composition of structural units by fusing their transitions tf(_, ..., _).

For Petri nets providing for the possibility of multiple arcs, it is reasonable to introduce another *n*-ary "mixed" *whole-fusion* operator: wf(_, ..., _), that is the fusion operator of both positions and transitions of the same structural unit. Such *n*-fold self-fusing leads to a corresponding increase in the multiplicity of arcs *n* times while maintaining the original position and transition.

We will denote "C" as p-unit (O→|) and "I" as t-unit (|→O).

For functional Petri nets [10], we introduce the index form of the notation: position and transition numbers by lower indices; marking of positions (*m*), speeds of transitions (*v*) by upper indices; thresholds of arcs (*k*) and transition delays (*d*) (if necessary) by brace parametric representation. In this case, it is advisable to introduce end-to-end numbering of positions and transitions for a more compact PN representation (Fig.1).

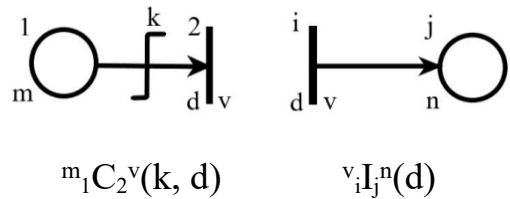

$${}^m_1 C_2^v(k, d) \qquad {}^v_i I_j^n(d)$$

Fig.1. Graphical representation and notation of p-unit and t-unit

We define the following rules for expressing a Petri net by formulas.
**R1**. Structural units are combined in the formula only by matching post-pre-indices.
**R2**. The common index for structural units can be bracketed: the pre-index is before the bracket, the post-index is after the bracket.

In PN-expression formula, the space remaining outside the index bracket will be denoted by empty place, for example, $_aC_k \circ {_bC_k} = (_aC, {_bC})_k$, where $\circ$ is the sign of both the units' composition before the fusion.

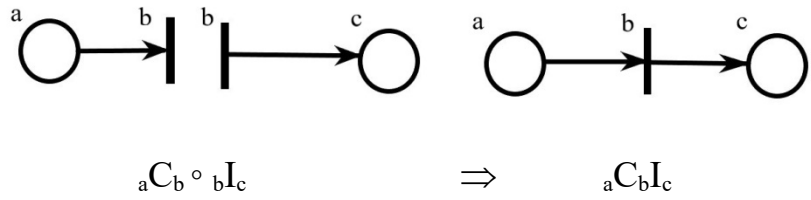

$_aC_b \circ {_bI_c} \quad \Rightarrow \quad {_aC_bI_c}$

Fig.2. The Fusion principle (rule R1)

Besides, for functional Petri nets with inhibitory and associative arcs [10], we introduce additional notation for the corresponding p-units: "B" for the unit with inhibitory arc (---|) and "A" for the unit with associative arc (- - ->). All the above presentation rules are fully true for them (Fig.3).

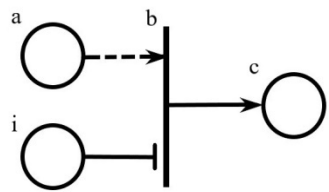

$_aA_b \circ {_iB_b} \circ {_bI_c} \Rightarrow (_aA, {_iB})_bI_c$

Fig.3. The result of fusion (rule R2)

**Proposition 3.** *Defusion Principle*. Any Petri net can be subjected to complete or partial decomposition into its structural units by applying the *defusion operator*: Df (*expression*) to the entire PN or any of its fragment.
For example,

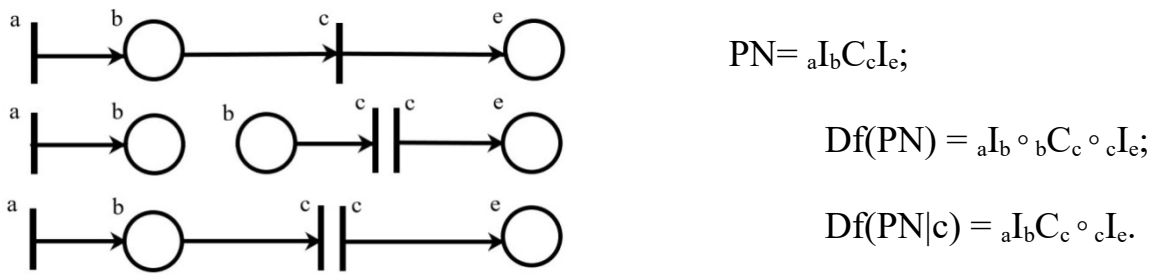

$PN = {_aI_bC_cI_e};$

$Df(PN) = {_aI_b} \circ {_bC_c} \circ {_cI_e};$

$Df(PN|c) = {_aI_bC_c} \circ {_cI_e}.$

Fig.4. Variants of defusion

Since the use of the defusion operator is a preparatory step for structural transformations in the PN, the net is not performed in the defusion step.

## 3. Petri Nets' Receptors

Nowadays, hybrids PN are commonly used in a wide range of areas and they are considered to bring maximum effect for modeling. That is why we have decided to describe some principles of building parameters' receptors, adapted namely for hybrids PN. We propose the concept of position marking receptor and transition velocity receptor.

a) Position marking receptor.

The purpose of the receptor is to 'feel' and 'show' the number of tokens, set in a position, connected to the receptor. According to the denotation principles, shown in [2], we will denote the position marking receptor as iRM, where 'R' means 'receptor', 'M' means 'marking' and 'i' - is a number of a position to be connected to the receptor. Its structure before the fusion is shown in Fig.5.

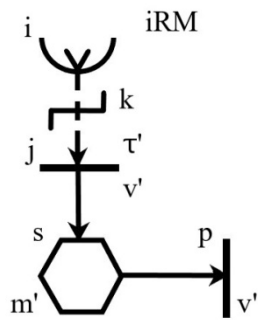

Fig.5. Position marking receptor

The simple receptor consists of three units in series. The first one, A, is p-fusing with the C-unit of the main part of PN and the collaboration of the first and the second (AI) units transmit the information from the position, without the resource consumption, to the "special" receptor's position ( ⬡ ), which shows the number of tokens in our estimated position on the previous step. The third unit stands for cleaning-up receptor's position in order to get new information on each step. If it is not required and we need to see a total number of tokens during the PN's work, we can switch the transition off, varying its parameters of velocity (V=0).

b) We will denote the transition velocity receptor as jRV, where 'V' means 'velocity'.

It's principle of work is very similar to the position marking receptor's. The difference lies in evaluating the transition's speed. The structure of the receptor is represented in Fig.6.

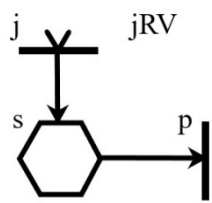

Fig.6. Transition velocity receptor

Receptor includes only two units: the first one, I, is t-fusing with the estimated transition from the main PN and assign the velocity value to the receptor's position, as its own marking. The second unit, C, works the same way as the third one in the position marking receptor and clean the receptor's position, taking tokens away via the outlet transition.
The "connection" of the receptor (effector) to the corresponding element of the main Petri net is carried out based on the Fusion principle (Fig. 7), whereupon the formed complex functions as a unified network.

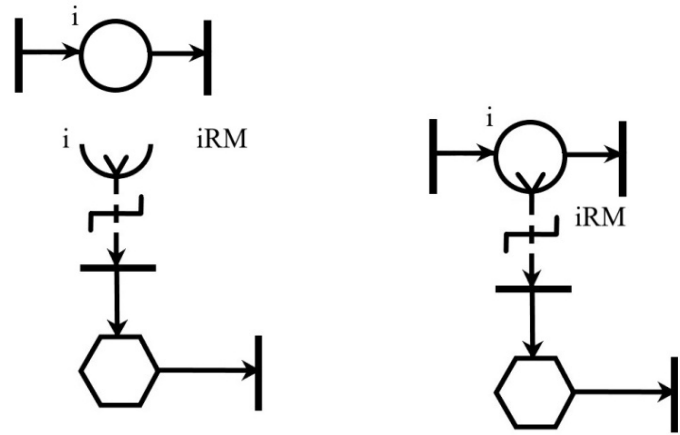

Fig.7. Implementation of the receptor to the main net's position

When there is a necessity to "disconnect" several receptors, this process is realized on the basis of the Defusion principle [2].

## 4. Petri Nets' Effectors

A system approach determines the presence not only receptors, but also effectors in a holistic net management structure.
Based on the initial data of the receptors, the *Solver* processes the information, and the effectors implement the specified controling actions in accordance with the decision. It is implemented by adding tokens or changing speeds in the elements of the main net.
We distinguish two types of effectors: position marking and transition velocity.
a) The sense of position marking effector is changing the marking value of the affected position, connected with the effector. We will denote it as EMi, where 'E' means 'effector', 'M' means 'marking' and 'i' - is a number of an effector affected position. Its structure before the fusion with the position is shown in Fig.8.

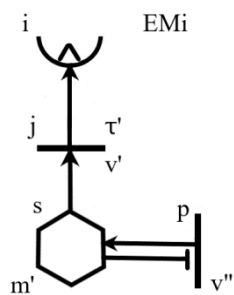

Fig.8. Position marking effector

The simple effector is built of three units, but unlike the receptor's formation we will describe the effector's formation from the lower unit. The first unit, I, provides the resource flow to the effector's position. The number of transferred tokens is set via the input transition's parameters. The second and the third units (CI) transfer tokens to the main net's position with the adjusted capacity.

b) The transition velocity effector is denoted as EVi, where 'V' means 'velocity'. Its sense is to transmit the required value of velocity to the main net transition (Fig.9).

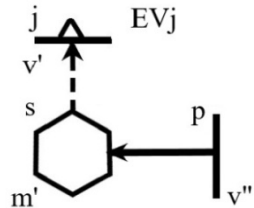

Fig.9. Transition velocity effector

It consists of two units, where the first unit (I) supplies the effector's position with the needed number of tokens, which corresponds to the required value of velocity. The second one (C) transfers it to the main net's transition, to be connected with the effector.

It should be noted that in addition to the ordinary (single, simple) receptors and effectors, their complex varieties are possible, in principle. For example, there may be a complex position marking receptor (Fig. 10), which allows to obtain a total number of tokens in positions a,..., b of the main Petri net at each step.

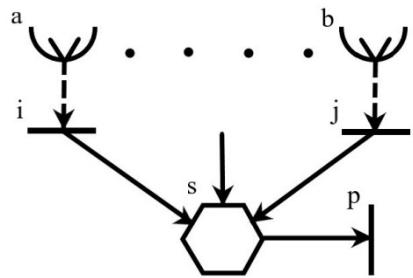

$$_xRM, x = \{a,\ldots, b\}$$
$$_xRM = (_aA_iI,\ldots, _bA_jI)_sC_p$$

Fig.10. Complex position marking receptor

In turn, the complex position marking effector (Fig. 11) allows to distribute a given number of tokens to positions a, ..., b of the main Petri net. In general case: $V_a \neq V_b, \forall a, b \in x$.

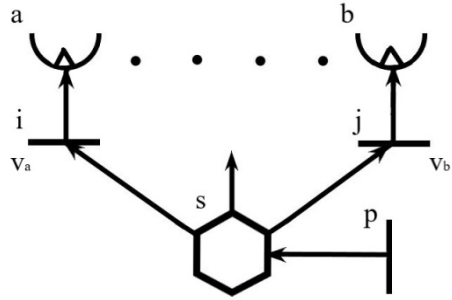

$$EM_x, x = \{a,\ldots,b\}$$
$$EM_x = {}_pI_s(C_iI_a,\ldots,C_jI_b)$$

Fig.11. Complex position marking effector

The physical meaning of the parameters of the above-described PN-receptors and PN-effectors is given in the table.

|  | iRM | iRV | EMi | EVi |
|---|---|---|---|---|
| m' | Perceived marking value | Perceived velocity value | The number of transferred tokens | The value of transferred velocity |
| v' | Capacity | - | Capacity | - |
| v'' | The number of filling/reset cycles | The number of filling/reset cycles | Token feed rate | The formation of the required velocity |
| τ' | Receptor's response rate | - | Effector's response rate | - |
| k | Receptor's response threshold | - | - | - |

The capacity of iRM and EMi can be determined by two factors: the receptor's (effector's) intrinsic capacity and the externally specified values. In the first case, the readout number of tokens may appear higher than the capacity and then the readout is carrying out in several steps. In the second case (V '= m'), the capacity is considered unlimited.

5. **Applications**

Receptors and effectors can be used both individually (independently) and as a part of an integrated control system/solver. An example of a Petri net, which is supplied with all of four types of receptors/effectors, is presented in Fig. 12.

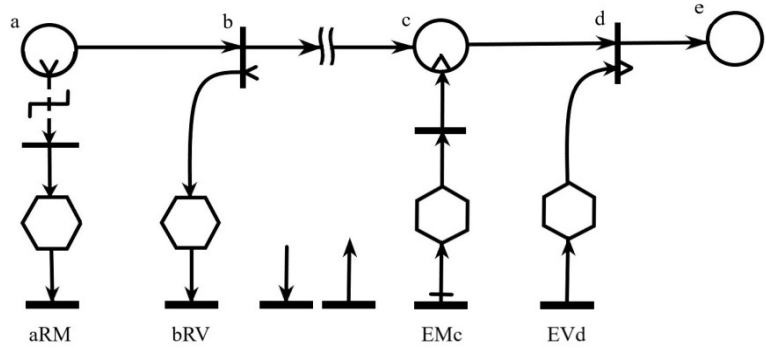

Fig.12.

For the teamwork of receptors and effectors, we need to introduce one more complex structure, which we call *Solver* (Fig.13). The Solver analyzes received information from the PN-receptors and determines the control parameters implemented by PN-effectors.

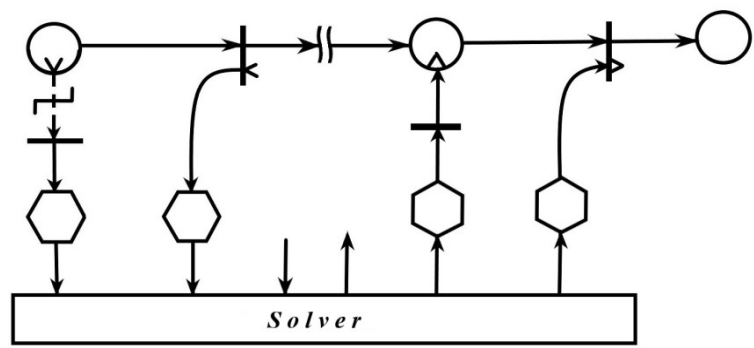

Fig.13. Complex Petri net structure

**Example**. For the manufacture (3) of some products (4), a resource (2), which is produced from (1), is required. The delivery of these products to the store (6) occurs at a certain speed (5). There is a certain number of customers (8), to whom these products are supplied with the intensity (7).

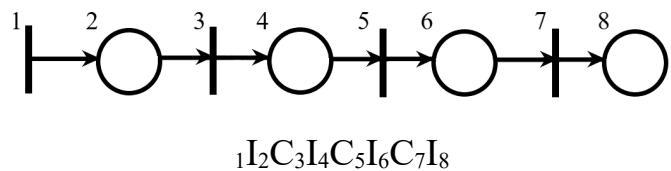

$_1I_2C_3I_4C_5I_6C_7I_8$

Fig.14. Main Petri net and its formula

The renewal of the resource (2) is carried out with a certain speed (1), ultimately insufficient to satisfy the needs of all customers (8). Therefore, to determine the amount of the resource in position 2, we set the position marking receptor (2-9-10-11) using the fusion principle. The Solver (12-13-14) processes this information and determines how much resource needs to be added to (2), which is implemented by the position marking effector (15-16-2). At each step, the information read by the receptor (2-9-10-11) is updated, and the previous value is taken away via the outlet transition (11).

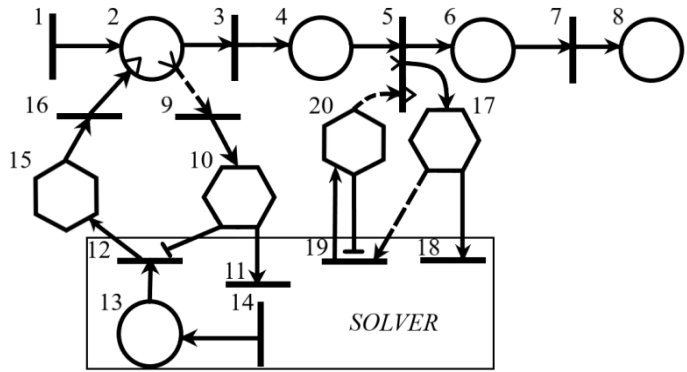

$_1I_2(A_9I_{10}(C_{11},B,_{14}I_{13}C)_{12}I_{15}C_{16}I)_2C_3I_4C_5(I_{17}(C_{18},A,_{20}B)_{19}I_{20}A)_5I_6C_7I_8$

Fig.15. Complex Petri net and its formula

Since, as a result of regulation of marking (2), the quantity of products (4) delivered to the store (6) also changes, it is necessary to adjust the speed of delivery (5). To do this, the transition velocity receptor (5-17-18) is set on transition (5). The *Solver* analyzes the speed value at the transition (5) and adds to (5) the missing speed value via the effector (20-5). At each step, the speed value obtained by the receptor (5-17-18) at the previous measure is taken away via the outlet transition (18). This scheme is adaptive to the varying of the number of customers and intensity of consumption.

The above model can be implemented in modeling environments (tools) oriented to functional Petri nets, in particular, in Cell Illustrator [10]. An example of such an implementation for 15 consumers is shown in Fig. 16.

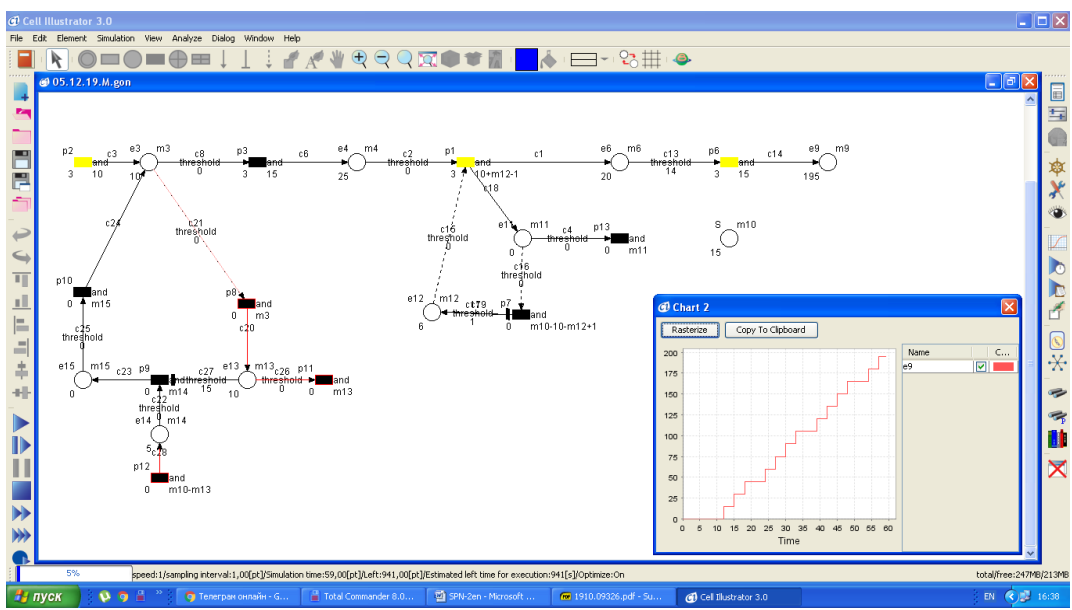

Fig.16. Simulation result in the Cell Illustrator tool

## 6. Conclusion

Receptors (detectors, sensors) and effectors (actuators, drives, execution units) are an integral part of most control systems, complex systems, and AI systems. It seems natural to include them in the model component of such a powerful modern modeling apparatus as Petri nets.

Is it possible to implement receptors/effectors in Petri nets in the formal concept of Petri nets? This paper answers this question.

We proposed the concept of PN-receptors and PN-effectors for high-level nets, namely for functional Petri nets. In this paper, we consider only two, in our opinion, the most significant types of receptors/effectors: the position marking and the transition velocity (rate). The remaining parameters of the Petri nets are still awaiting their "receptor/effector" embodiment.

Applying the Fusion/Defusion principles makes possible PN-modeling of "flexible" systems, in which receptors and effectors can be not only components of the system, but can be connected from the outside at a given time, for a given interval, and navigate through the structural elements of the system etc. The proposed concept enables to expand the capabilities of traditional Petri nets for adequate modelling of a wide class of systems.